\documentclass[%
 aip,
 amsmath,amssymb,
preprint,%
]{revtex4-1}

\usepackage{graphicx}
\usepackage{dcolumn}
\usepackage{bm}

\usepackage[utf8]{inputenc}
\usepackage[T1]{fontenc}
\usepackage{mathptmx}

\begin{document}

\preprint{AIP/123-QED}

\title[New Emission Algorithm]{A new simple algorithm for space charge limited emission}

\author{P. H. Stoltz}
%
\email{pstoltz@txcorp.com}
\affiliation{ 
Tech-X Corporation
}%

\author{J. W. Luginsland}
\affiliation{%
Confluent Sciences, LLC.
}%

\author{A. Chap}
\author{D. N. Smithe}
\author{J. R. Cary}
\affiliation{%
Tech-X Corporation
}%

\date{\today}

\begin{abstract}
Many high power electronic devices operate in a regime where the current they draw is limited by the self-fields of the particles.  This space-charge-limited current poses particular challenges for numerical modeling where common techniques like over-emission or Gauss’ Law are computationally inefficient or produce nonphysical effects.  In this paper we show an algorithm using the value of the electric field in front of the surface instead of attempting to zero the field at the surface, making the algorithm particularly well suited to both electromagnetic and parallel implementations of the PIC algorithm.  We show how the algorithm is self-consistent within the framework of finite difference (for both electrostatics and electromagnetics).  We show several 1D and 2D benchmarks against both theory and previous computational results.  Finally we show application in 3D to high power microwave generation in a 13 GHz magnetically insulated line oscillator.  
\end{abstract}

\maketitle

\section{\label{sec:Introduction}Introduction}

High power microwave (HPM) devices frequently operate at the highest possible voltage in order to produce the maximum rf power.  The maximum power typically is coincident with the maximum current, and frequently the maximum current is limited by the self fields (or space charge) of the emitted electrons.  This situation serves to maximize the input power to the high power electromagnetic source.  This space charge limiting (SCL) current is well understood both theoretically and experimentally, however computer models can struggle to reproduce the exact physical limiting current.  This stems primarily from the formal singularity that exists for the space charge limiting current with electrons emitted with zero emission velocity~\cite{Zhang17}.  One computational approach is to over-emit electrons and let the self-consistent electric field push the excess charge back to the cathode.  This approach has a number of disadvantages, including being computationally inefficient and resulting in a nonphysical virtual cathode that must be resolved by the mesh at the point where the electrons are pushed back. Another popular approach is to use Gauss’ Law to only emit enough charge to exactly zero out the electric field at the emission surface.  While this is an improvement on over-emission, the Gauss’ Law approach is still quite inefficient computationally.  For instance, this approach requires either a particle sort and potentially an additional electrostatic solve.  Especially in electromagnetic applications, the extra sort and solve are burdensome.  Furthermore, in a parallel particle-in-cell implementation, both the sort and electrostatic solve can be less efficient.  The goal of this work is to find a new approach to SCL emission that is both physically accurate and computationally efficient. 

In this paper we show an algorithm using the value of the electric field in front of the surface instead of attempting to zero the field at the surface.  Using the well-established particle-in-cell (PIC) code VSim~\cite{vorpal04,MNL09}, we show how the algorithm is self-consistent within the framework of finite difference (for both electrostatics and electromagnetics).  We show several 1D and 2D benchmarks against both theory and previous computational results for the electrostatic case.  We also show a benchmark 1D electromagnetic case, showing that the voltage rise time can impact transient behavior in a manner consistent with the literature on electromagnetic effects on space-charge limited flow.  Finally we show application in 3D to high power microwave generation in a 13 GHz magnetically insulated line oscillator. 

\section{\label{sec:Derivation}Derivation of the algorithm}

\begin{figure}
\includegraphics[width=2.66in]{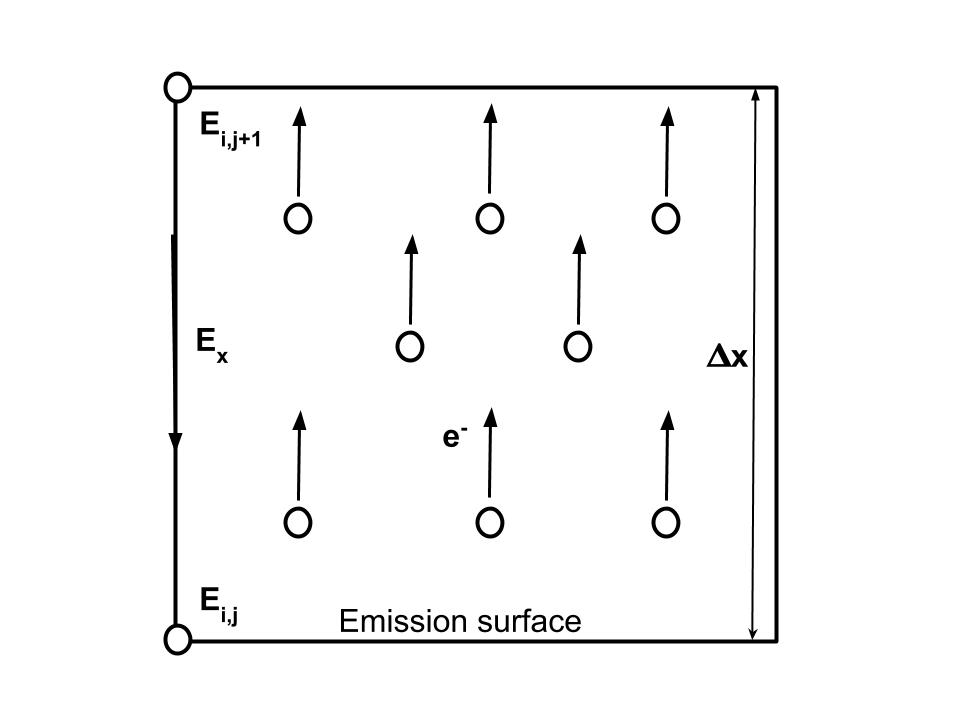}
\caption{\label{fig:FDTDGrid} A typical finite difference scheme, showing the electric field ($E_x$) one half of a cell offset from the surface and the interpolated nodal fields ($E_{i,j}$ and others).  An electrostatic scheme would include charge density on the cell corners.  An electromagnetic scheme would include current density co-located with the edge electric field.}
\end{figure}

The motivation for this algorithm is to use the finite difference electric field just above a surface to inform the choice of emitted current.  Previous techniques typically relied instead on trying to interpolate the field to the surface, inject charge and then use Gauss’ Law to zero out the surface field.  
However, we assume that the finite difference electric field at the midpoint of the cell edges (see $E_x$ in Fig.~\ref{fig:FDTDGrid}) is the field resulting from space charge limited emission, and we use that fact to inform our choice of particle velocities (or equivalently, current densities).

In particular, we make the assumption that near the emission surface, the system is emitting steady-state space charge limited current.  For a one-dimensional, Cartesian geometry, the steady-state limiting current in terms of the electric field and cell size is given by the Child-Langmuir law:
\begin{equation}
J_{SCL} = \frac{4}{9}\epsilon_0\sqrt{\frac{2q}{m}}\frac{E^{3/2}}{\Delta x^{1/2}}.
\label{eq:one}
\end{equation}
In the above, $\Delta x$ is the cell size and $E$ is the normal component of the electric field interpolated to the position of a new particle being added as part of the emission.  In VSim, this interpolation involves first interpolating the edge fields to the nodes (labeled $E_{i,j}$, etc, in Fig.~\ref{fig:FDTDGrid}), and then interpolation from the nodal fields to the particle position.  By using the same interpolation scheme for evaluation of the space charge limited emission as we do for the particle push, we are building in a self-consistency to the scheme.  We believe this consistency is one reason for the performance of this algorithm despite the potential challenge of applying the Child-Langmuir solution to just the first cell rather than the more traditional procedure of applying the law to the entire gap~\cite{Luginsland_review}.

We can rewrite Eq.~\ref{eq:one} in several useful ways, specifically we can write the density and velocity as a function of position~\cite{Umstattd_AJP}:
\begin{equation}
\rho(x) = \left(\frac{2}{9}\frac{m\epsilon_0J_{SCL}^2}{q}\right)^{1/3} x^{-2/3}.
\label{eq:two}
\end{equation}
\begin{equation}
v(x) = \left(\frac{9}{2}\frac{qJ_{SCL}}{m\epsilon_0}\right)^{1/3} x^{2/3}.
\label{eq:three}
\end{equation}
This scaling of density and velocity with position will be important when we examine the convergence of our algorithm.  While Eq.~\ref{eq:one} paradoxically appears to depend on the cell size, the electric field in the 1D Child-Langmuir law scales as~\cite{Umstattd_AJP} $x^{1/3}$, so $J_{SCL}$ remains constant as the cell size decreases.  

For the cases shown here, we use a variable weight approach, with a fixed number of emitted computational particles per step, placed randomly within the cell, and the weight of the particles varied to match the current density given by Eqs.~\ref{eq:one}. We give a list of the steps we follow here:
\begin{enumerate}
  \item Pick a number of particles to emit per step (usually based on computational resources and accuracy requirements)
  \item For each particle, pick a location within one cell of the emission surface
  \item Evaluate the electric field at the particle location, using the same interpolation as for the force on the particle
  \item Take the component of electric field normal to the emission surface and use that normal electric field and the cell size in Eq.~\ref{eq:one} to get a $J_{SCL}$
  \item Multiply that $J_{SCL}$ by the time step and the emission area, and divide it by the number of particles from Step 1, to obtain the weight for this particle.
\end{enumerate}
For a fixed-weight approach, one would add particles with a probability scaled by the density profile, keeping track of the current added, until the space charge limit is reached. For our implementation, we added all particles with zero initial velocity, however one could use Eq.~\ref{eq:three} to obtain potentially even more self-consistency.  The advantages of this approach are that it requires no extrapolation of the electric field to the surface, requires no additional evaluation of Gauss' Law and no associated sorting of particles, and does not rely on numerical virtual cathode formation.  Thus, the method is both easy to implement in parallel PIC codes and highly effective as we show below.

\section{Verification and Validation of 1D Electrostatics}
\begin{figure}
\includegraphics{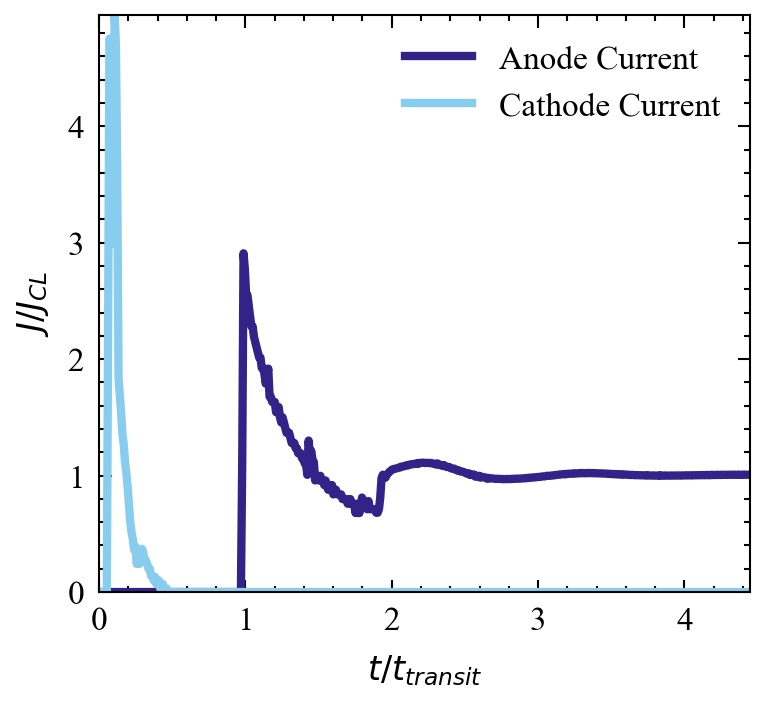}
\caption{\label{fig:1DESCurrent} The anode (dark blue) and cathode (light blue) return current versus time in a 1D planar diode simulated with 2000 cells.  The new algorithm results in steady current at the expected theoretical space charge value after about two transit times after current arrives at the anode.}
\end{figure}

\begin{figure}
\includegraphics{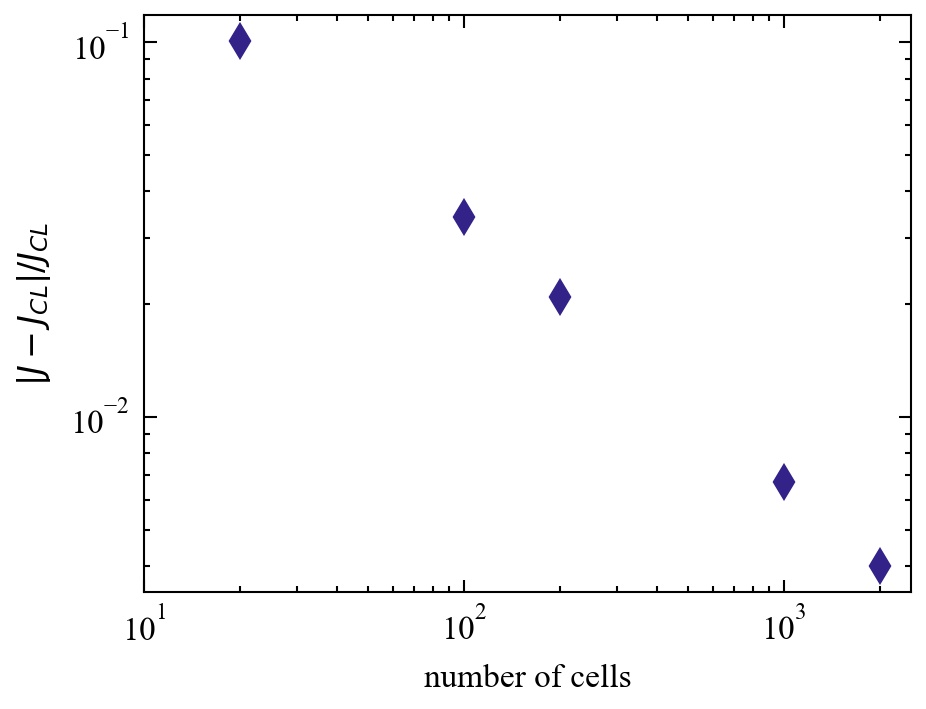}
\caption{\label{fig:1DESConvergence} The convergence of the current to the theoretical value with number of cells.  The error converges like the number of cells to the -2/3 power, consistent with the fact the theoretical electron density scales as 2/3 power.}
\end{figure}

\begin{figure}
\includegraphics{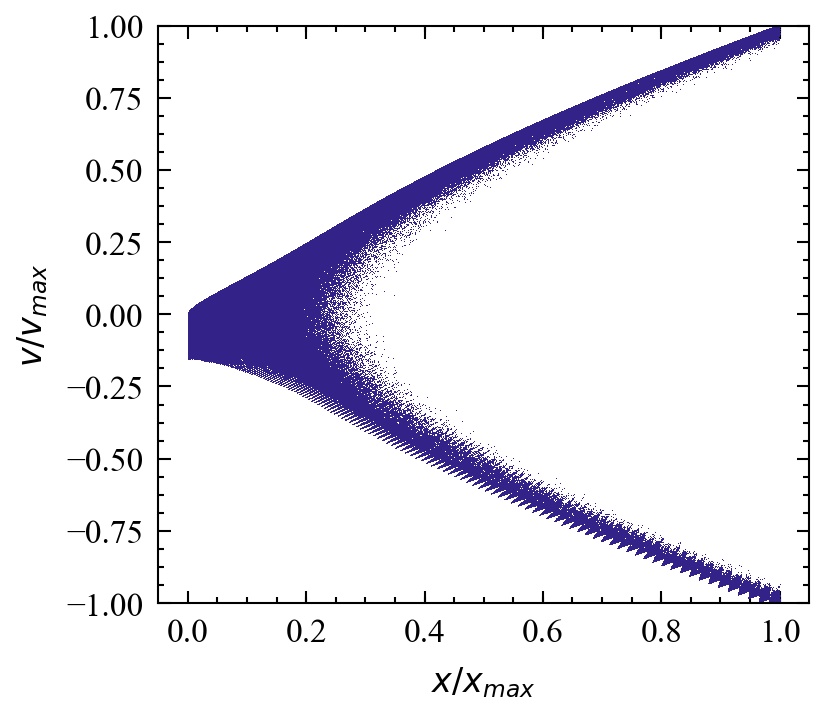}
\caption{\label{fig:1DESReflectionPhaseSpace} The electron phase space for the case where the anode is a perfect elastic electron reflector.  This case shows that the new algorithm works not just for simple cases, but also for cases where charge or current already resides in a cell.}
\end{figure}

\begin{figure}
\includegraphics{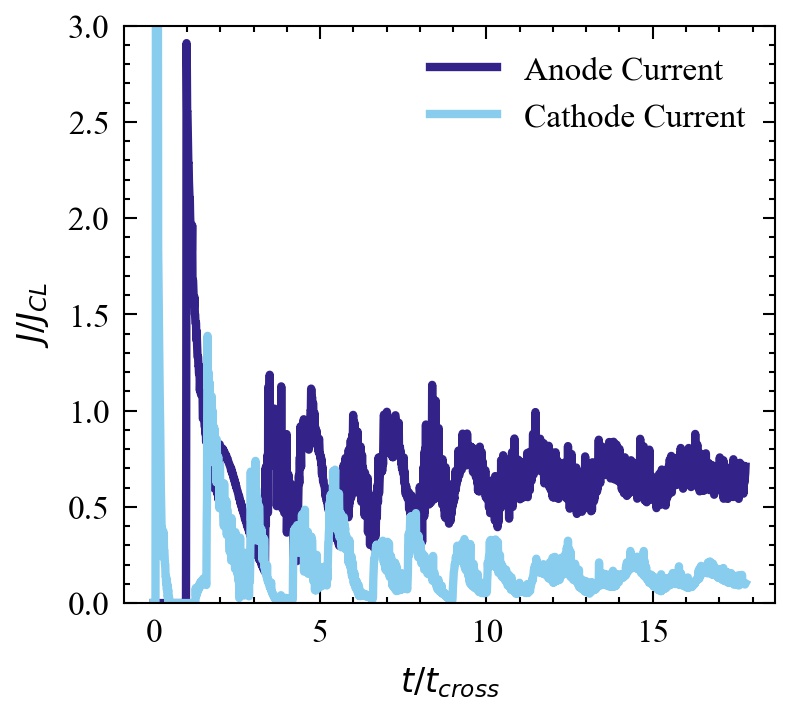}
\caption{\label{fig:1DESReflectionCurrent} The current density at the anode (dark blue) and re-absorbed at the cathode (light blue) for the case of an anode that emits elastic secondary electrons.  The simple estimate of steady-state current at the anode of 0.5 $J_{SCL}$ is a slight under-estimate due to the finite transit time that allows a virtual cathode to form, resulting in some return current to the cathode.}
\end{figure}

The initial test case for our new emission algorithm is a 1D planar diode.  We chose unit voltage and gap distance for this test, and we chose computational parameters that resulted in roughly 100,000 simulation particles in the steady state.  We show in Fig.~\ref{fig:1DESCurrent} the current density versus time at both the anode and cathode.  The current density is normalized to the theoretical 1D limit, and the time is normalized to the time for an electron to cross the gap in the absence of space charge.  For the simulations in this figure, we used 2000 cells across the diode gap.  The new algorithm exhibits a transient period where current does return to the cathode, but on a time scale short compared to the vacuum gap transit time, the cathode current drops to zero and remains zero, demonstrating that this new algorithm results in no numerical virtual cathode formation.  Furthermore, by three transit times, the anode current density is steady at the expected space charge limit.
We show the scaling of the error with resolution in Fig.~\ref{fig:1DESConvergence}.  For this plot, we held the time step and number of particles in the simulation fixed, and we varied only the number of cells.  We found that error scaled like the number of cells to the -2/3 power.  This is consistent with the fact that the theoretical Child-Langmuir electron density scales as $x^{-2/3}$, as shown in Eq. 2, indicating that the current is only as accurate as the simulated density near the cathode.
As one further test of the 1D electrostatic behavior, we tested how this algorithm behaves in the presence of particles moving in the opposite direction, for example due to elastic reflection of secondary electrons.  We wanted to test how rapidly the algorithm can adapt to this non-standard case.  We chose as a simple test to reflect elastically all electrons at the anode.  In this case, the charge at each position is roughly doubled (the particles moving toward the anode and the particles moving toward the cathode), so one expects the steady-state cathode-to-anode current to be roughly half the Child-Langmuir value.  In Fig.~\ref{fig:1DESReflectionPhaseSpace}, we show the longitudinal phase space for this case after roughly ten crossing times, showing the presence of the return current. In  Fig.~\ref{fig:1DESReflectionCurrent}, we show the current reaching the anode in dark blue and the current returning to the cathode in light blue.  The rough estimate of 0.5 $J_{SCL}$ is slightly under the simulated anode current in the steady state.  This is due to the finite transit time across the gap and back, during which a virtual cathode forms (seen in Fig.~\ref{fig:1DESReflectionPhaseSpace}).  This results in return current to the physical cathode.  The amount of this return current is the amount by which the anode current exceeds 0.5 $J_{SCL}$.  This test demonstrates that the algorithm handles not just pure vacuum cases, but also cases where charge or current (even oppositely directed) is already present in a computational cell. 
\section{Verification and Validation of 2D Electrostatics}
\begin{figure}
\includegraphics{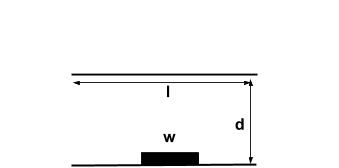}
\caption{\label{fig:2DESGeom} The geometry for the finite width 2D test case.}
\end{figure}

\begin{figure}
\includegraphics{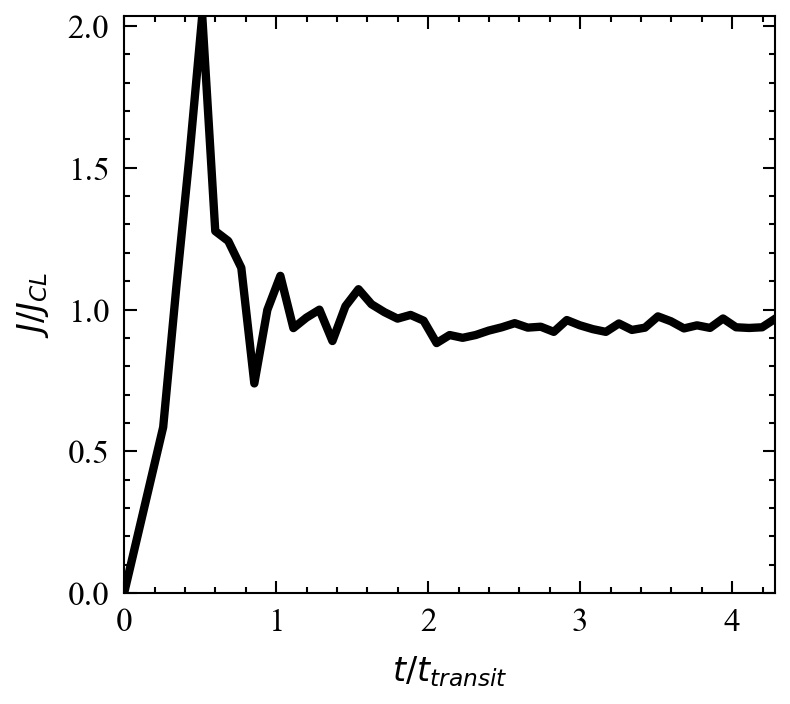}
\caption{\label{fig:2DESCurrentvT} The current density at the center of the emitter versus time.  The algorithm we have implemented takes only a few transit times to find a steady state that has less than 10\% variation from the expected value.}
\end{figure}

\begin{figure}
\includegraphics{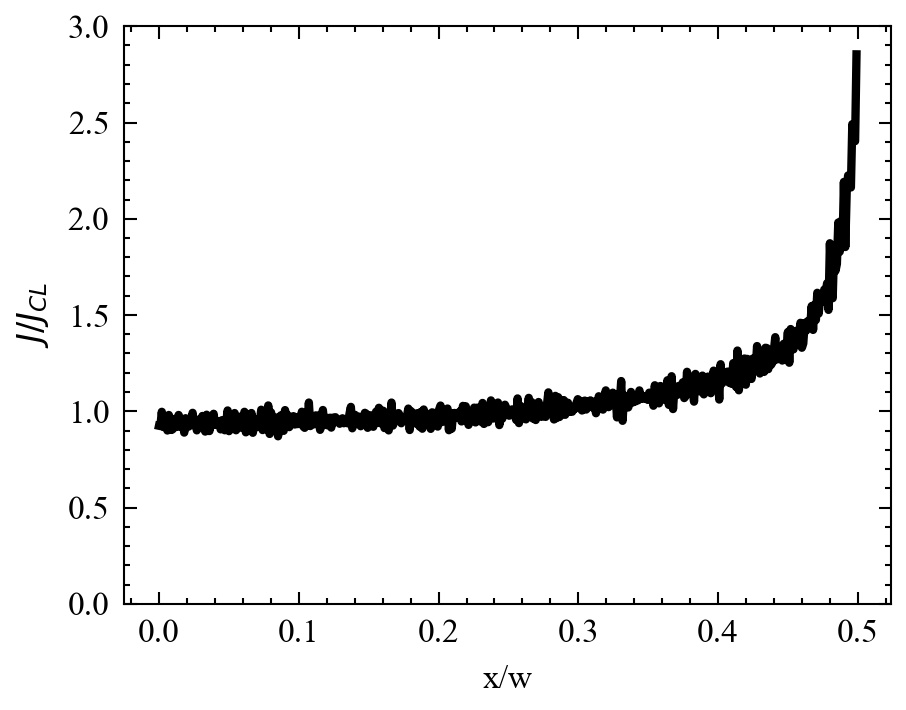}
\caption{\label{fig:2DESCurrentvX} The current density as a function of space after several transit times for a 2D planar diode with a finite emission region of width w.  The factor of three enhancement matches previous results, further demonstrating that this new algorithm is working properly.}
\end{figure}

A second benchmark test with widely accepted results is a 2D planar diode with a finite width emission region~\cite{Watrous01,WLF01}.  This case illustrates one key difference from the 1D diode, namely the appearance of enhanced current density at the edges of the emission region.  The exact amount of enhancement depends on the width of the emission region, the size of the diode gap, and the strength of any focusing magnetic field, but is on the order of three times the 1D Child-Langmuir value for emission width and diode gap close to the same size with a strong focusing field.
Therefore, we set up in a 2D planar diode with dimensions (following previous work~\cite{Watrous01}), d=1cm, w=1cm, l=8cm, as shown in Fig.~\ref{fig:2DESGeom}.  Further following previous work, we simulated V=30kV across the gap, d, and applied a static magnetic field, By = 0.5T to keep the flow essentially purely in the y-direction (this field is roughly 10x larger than the field required to keep the Larmor radius under 1cm, even at the largest electron velocities).  For these parameters, the transit time is on the order of 0.2ns.
We ran our simulation for several transit times (1.6 ns) to allow for an equilibrium to form.  We chose computational parameters such that there were roughly 100,000 simulation particles at steady state.  We selected electrons within 2mm vertically of the emission region and measured the current density, both as a function of time and space.  In Fig.~\ref{fig:2DESCurrentvT} we show the current density at the center of the emission region compared to the 1D Child-Langmuir versus time.  Because space charge limited flow is a steady state condition, one expects effects like the initial spike over time scales short compared to the transit time (this effect has a physical analog in the short pulse community, where researchers can generate short electron pulses with current density greater than the Child-Langmuir value if the pulse duration is shorter than the transit time~\cite{Koh06}).  After a few transit times, the current density is steady to within 10\% of the 1D Child-Langmuir value.
In Fig.~\ref{fig:2DESCurrentvX}, we have plotted the current density versus transverse position across the emission region.  The factor of three enhancement in the edge is a well-established characteristic of finite width 2D emission~\cite{Watrous01} and further establishes that this new algorithm is behaving as expected.
\section{Verification and Validation of 1D Electromagnetics}
\begin{figure}
\includegraphics{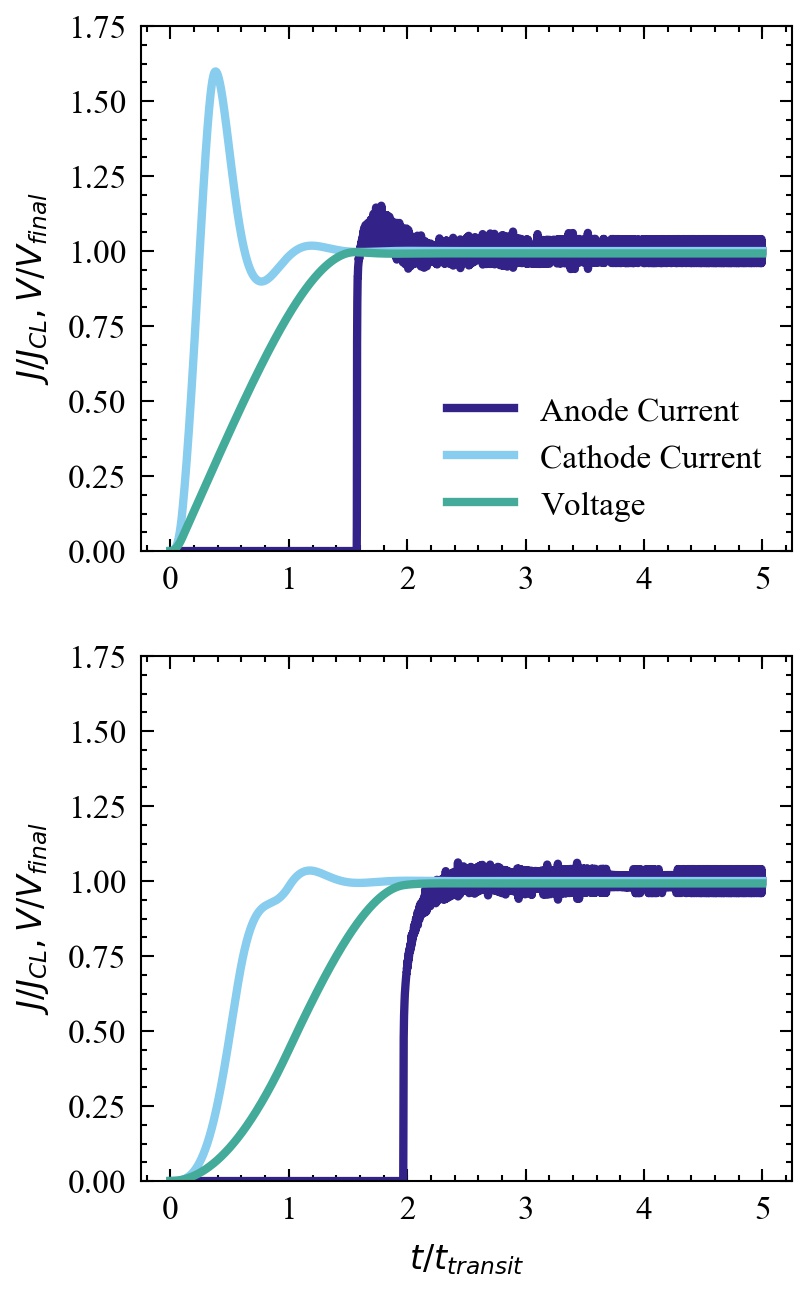}
\caption{\label{fig:1DEMCurrent} The voltage (green), emitted current density (light blue) and absorbed (dark blue) current density for an electromagnetic model of a 1D diode.  We show two voltage waveforms, a rapid linear rise (top) and a more gradual rise (bottom).  The new algorithm is able to capture both time-dependent (inductive spikes) and steady-state behavior.}
\end{figure}

The above tests demonstrated how the algorithm behaves for systems with time-independent fields, and in those tests we used an electrostatic field solver.  However, a main goal of this work is that this algorithm also performs well for electromagnetic systems.  Consequently, we repeated our test of a 1D diode using an electromagnetic field solver, and allowing for a transient voltage in the gap.  We changed the values of voltage and gap size to better suit an electromagnetic system, so we chose Vfinal = 50kV and d = 0.01m.  We chose computational parameters of 200 cells across the gap and roughly 100,000 particles in the simulation in the steady state.
To demonstrate the performance of the algorithm in a time-dependent setting, we performed two tests: one with the voltage varied quickly compared to the transit time, and one with the voltage varied slowly.  We show in Fig.~\ref{fig:1DEMCurrent} the results for the voltage (green), current density emitted from the cathode (light blue), and current density absorbed at the anode (dark blue).  In the first case (top), we varied the voltage linearly over one transit time.  In the second case (bottom), we varied the voltage more gradually over two transit times.  In both cases, the simulation reaches a steady state after 2.5-3 transit times, and in both cases, the simulation had zero return current to the cathode, indicating no numerical virtual cathode formation.  These metrics indicate that the algorithm performs well for electromagnetic simulations.  The algorithm also captures time-dependent transients such as the inductive increase in electric field for quick voltage rise, seen in the top plot as the spike of emitted current to 1.5 times the steady-state Child-Langmuir value at around 0.5 transit times.  The more gradual waveform used in the second case reduces the inductive effect.  This behavior is consistent with the physics of electromagnetic space-charge limited flow~\cite{EM_SCL}.

\section{Application to the physics of a MILO}
\begin{figure}
\includegraphics{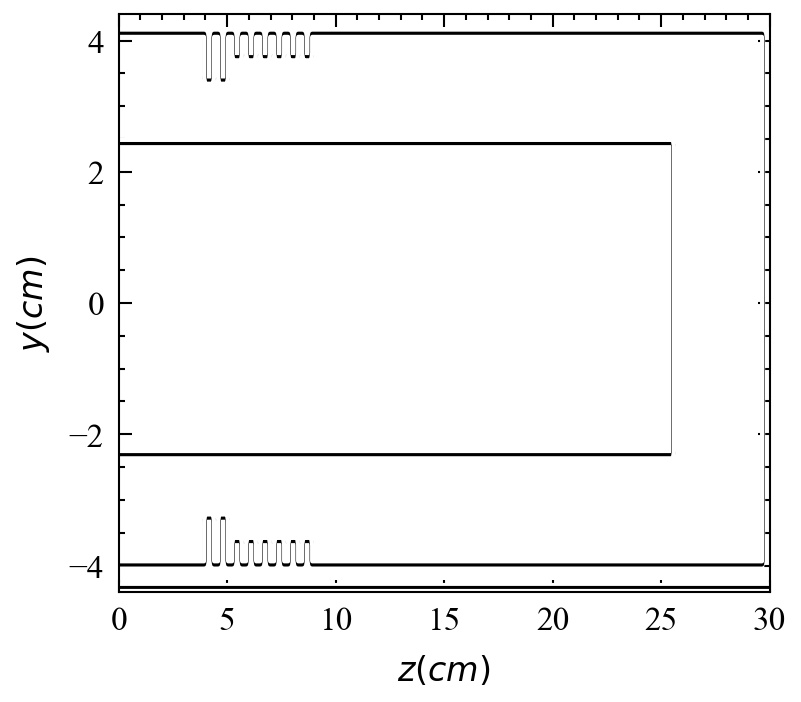}
\caption{\label{fig:MILOGeom} The cross sectional geometry of a 13 GHz MILO, showing the anode, cathode, and slow wave structure.  We monitor the voltage between the anode and cathode and in one of the cavities of the slow wave structure.}
\end{figure}

With the new algorithm well benchmarked, we apply it to a problem of interest in the HPM community, namely the physics of a magnetically insulated line oscillator (MILO), a promising HPM source for high frequency.  We chose to simulate a MILO similar to one from the literature~\cite{Jiang15} to allow us to compare our results with known values.  In simulation, these researchers saw 45 kA at 475 kV in the oscillator region, for an input power of just over 20 GW.  We show a cross section of the full 3D geometry in Fig.~\ref{fig:MILOGeom}.  For simplicity, we chose to omit the extractor and model only the oscillator region.  The slow wave structure has a choke region with two discs of slightly larger area, followed by six discs of smaller area.  An additional reason for choosing this particular geometry is the high frequency, which results in smaller bunching, and provides an even more strict test of the new emission algorithm.

\begin{figure}
\includegraphics[width=2.66in]{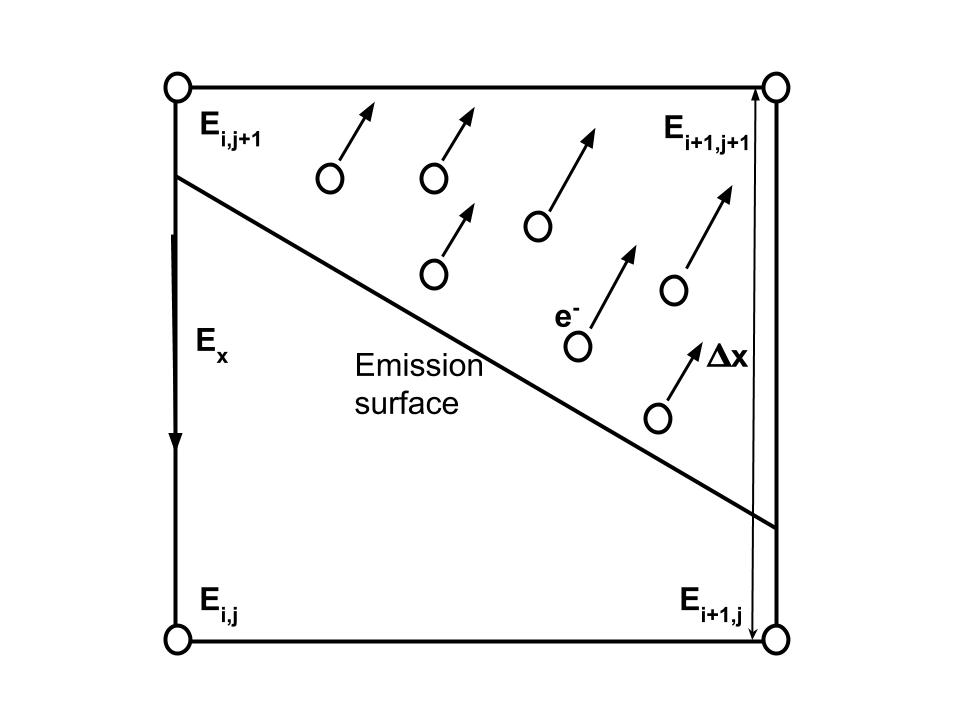}
\caption{\label{fig:CutCellGrid} A finite difference electromagnetic scheme with an embedded boundary.  Nodal fields within the metal ($E_{i,j}$ and $E_{i+1,j}$ in this example) will have in general smaller values than nodes in the vacuum, and this is how the space charge limited current then is modified in the cut cells.}
\end{figure}

The MILO is a 3D cylindrical structure that does not conform to the rectangular grids used in a standard finite difference scheme.  VSim uses an embedded boundary scheme to more accurately represent complex shapes within a rectangular grid.  As an example, we show in Fig.~\ref{fig:CutCellGrid} how a metal surface that does not conform to the grid lines might be represented.  
We have implemented an emission algorithm that follows these same embedded boundaries.  We use the same five-step algorithm shown in Sec.~\ref{sec:Derivation}, with the same approach of interpolating from the nodal fields for self-consistency (even if the nodes involved are within the metal).  In general, with the embedded boundaries, the nodal fields that are inside the metal (for example, the nodes labeled $E_{i,j}$ and $E_{i+1,j}$ in Fig.~\ref{fig:CutCellGrid}) will have values potentially different from their values in the absence of the embedded boundary~\cite{Meierbachtol15}.  This is because adjacent cells may not contribute to those nodes as they would in the case of no embedded boundaries (for example, one can imagine the cell below the one shown would also be within the metal and therefore would not have any current to contribute to the nodes $E_{i,j}$ and $E_{i+1,j}$).  These modifications to the nodal fields are the way the emission algorithm is modified for embedded boundary emission.

\begin{figure}
\includegraphics{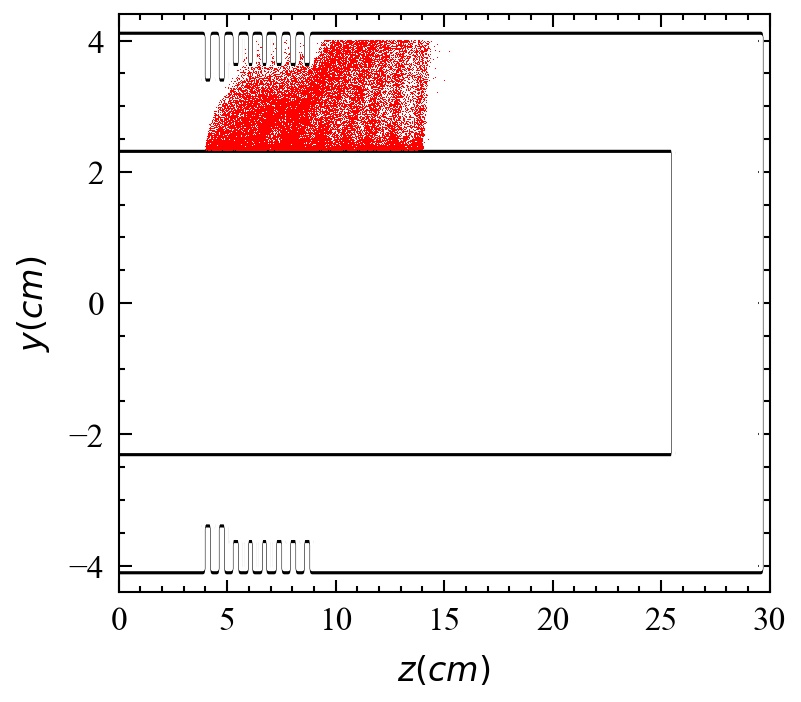}
\caption{\label{fig:MILOElecs} The simulated electron distribution at 25ns (showing the electrons only in the upper y plane for simplicity).  The strong bunching (seen in the vertical striations) is one sign of space charge limited emission working well.}
\end{figure}

\begin{figure}
\includegraphics{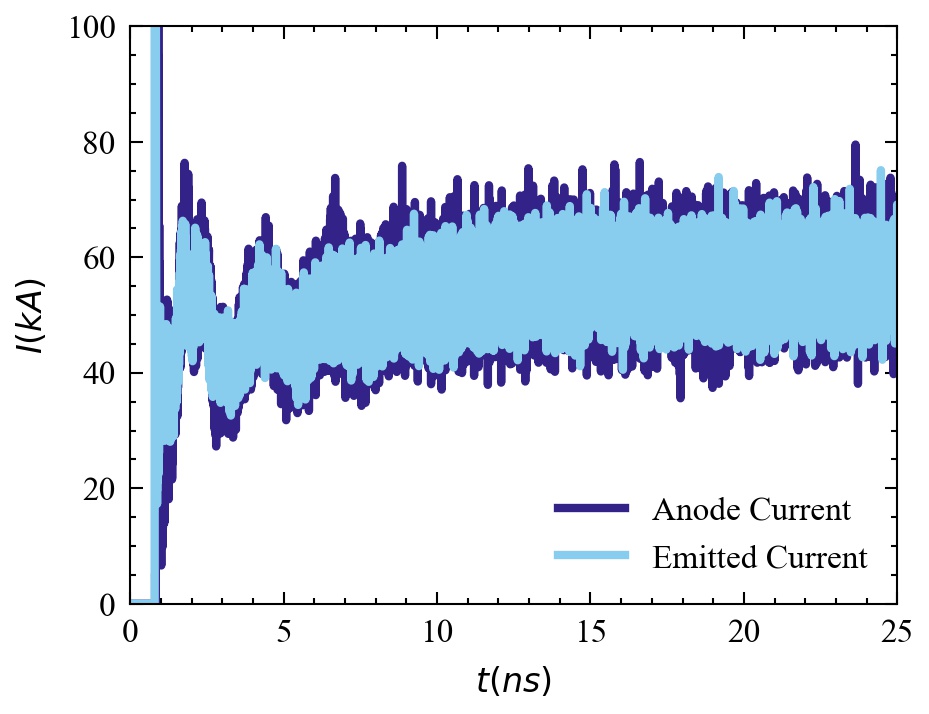}
\includegraphics{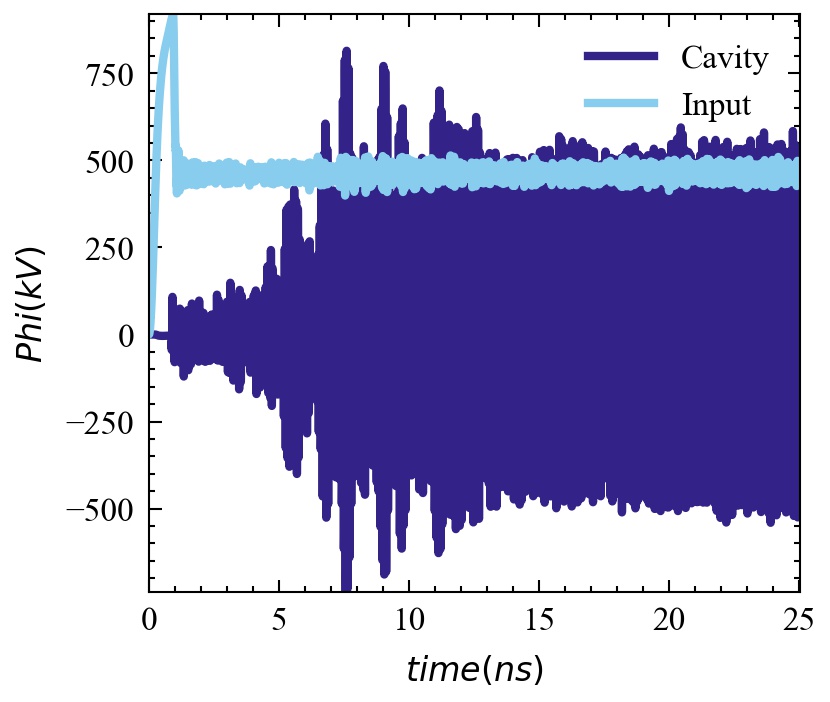}
\caption{\label{fig:MILOCurrentVoltage} The emitted current and current at the anode for the VSim model of the 13 GHz MILO (top), and the voltage input and measured in between one set of teeth in the slow wave structure (bottom).  The amplitude of the cavity voltage being equal to the input voltage is an indication that the space charge limited emission is working properly.}
\end{figure}

\begin{figure}
\includegraphics{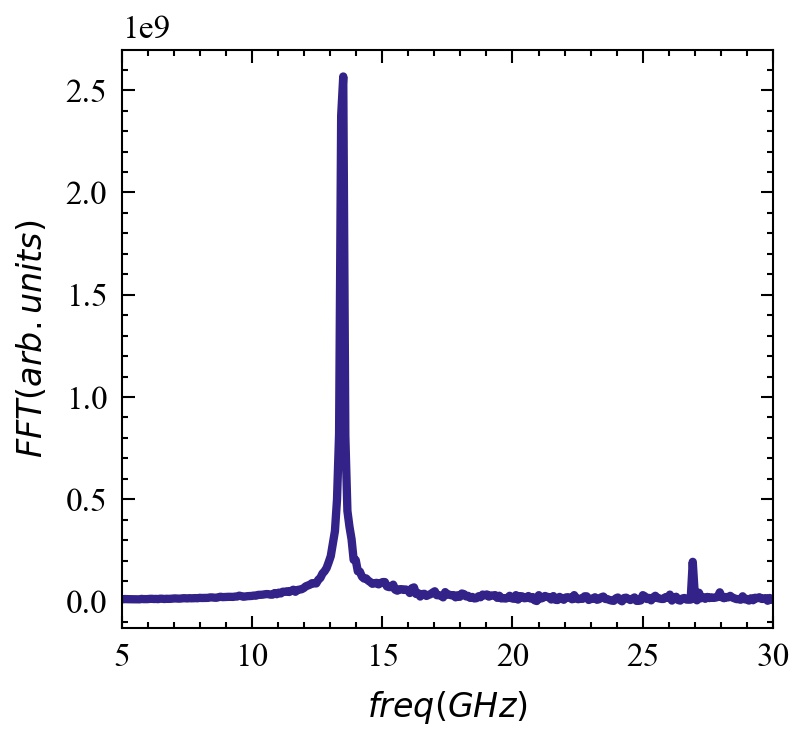}
\caption{\label{fig:MILOFFT} The FFT of the cavity voltage, showing primary signal at 13.5 GHz, near the 13 GHz value reported in the experiment.  The appearance of a two-times higher harmonic is an indication of strong coupling between the electrons and the fields, possible only when the space charge limited emission is working properly.}
\end{figure}

To isolate the performance of the emission algorithm, we proceed by removing the extractor from the MILO in the literature.  Without the extractor, in order to match the power of the real device, we chose a longitudinal cathode emission region to produce roughly the same amount of total current as seen in the previous simulations.  The SCL current per unit length (when corrected for the coaxial geometry and for relativistic effects~\cite{Zhang09}) for this geometry and voltage is roughly 500kA/m.  Therefore, we choose a longitudinal emission region of 0.1m (seen in Fig.~\ref{fig:MILOElecs}).  If the new algorithm is emitting at the SCL limit, we expect this to result in approximately 50kA.  Figure~\ref{fig:MILOCurrentVoltage} (top) shows that the current is approximately 55 kA, within 10\% of the expected value (because the geometry is not perfectly coaxial, we expect some difference).  The bottom plot in Fig.~\ref{fig:MILOCurrentVoltage} shows the applied voltage of 475kV and the rf cavity voltage with amplitude of nearly the same value.  This strong coupling of input voltage to cavity voltage is an indication of fully space charge limited flow.  Finally in Fig.~\ref{fig:MILOFFT}, we show the FFT of the rf cavity voltage, showing a strong peak at 13.5 GHz, within a few percent of the previously seen value of 13 GHz (the lack of the output coupler may explain this small difference).  Also in the FFT, one sees the second harmonic clearly present, another indication of the strong coupling between the electrons and the rf, possible only with space charge limited flow.   

\section{Conclusions}
In this paper we showed a new algorithm for simulating space charge limited emission.  The algorithm uses the value of the electric field in front of the surface instead of attempting to zero the field at the surface.  The algorithm uses the same interpolation scheme when evaluating the current as for evaluating the force on the particle, leading to a natural self-consistency.  We showed application of the algorithm to both electrostatic and electromagnetic test cases in 1D and 2D benchmarks against both theory and previous computational results.  We also showed application in 3D to high power microwave generation in a 13 GHz magnetically insulated line oscillator.  Finally, while we have implemented and tested this algorithm particularly for a PIC code, the same technique could work as a boundary condition for a fluid code.

\begin{acknowledgments}
This work was supported by the Office of Naval Research under Contract N68335-18-C-0060.  The authors thank Christine Roark for help with several details of the simulations.  The authors also thank Keith Cartwright, Kristian Beckwith, Thomas Gardiner, and Allen Robinson for helpful discussions.
\end{acknowledgments}

\pagebreak
\bibliography{aipsamp}
\pagebreak

\end{document}